# BIOLOGICAL EFFECTS OF PULSATING MAGNETIC FIELDS: ROLE OF SOLITONS

L. Brizhik[1,2]

[1]Bogolyubov Institute for Theoretical Physics, Metrolohichna Str., 14b, Kyiv 03680, Ukraine
[2]Wessex Institute of Technology, Ashurst, Southampton SO40 7AA, UK
E-mail: brizhik@bitp.kiev.ua

## ABSTRACT

In this paper, we analyze biological effects produced by magnetic fields in order to elucidate the physical mechanisms, which can produce them. We show that there is a chierarchy of such mechanisms and that the mutual interplay between them can result in the synergetic outcome. In particular, we analyze the biological effects of magnetic fields on soliton mediated charge transport in the redox processes in living organisms. Such solitons are described by nonlinear systems of equations and represent electrons that are self-trapped in alpha-helical polypeptides due to the moderately strong electron-lattice interaction. They represent a particular type of disssipativeless large polarons in low-dimensional systems. We show that the effective mass of solitons in the is different from the mass of free electrons, and that there is a resonant effect of the magnetic fields on the dynamics of solitons, and, hence, on charge transport that accompanies photosynthesis and respiration. These effects can result in non-thermal resonant effects of magnetic fields on redox processes in particular, and on the metabolism of the organism in general. This can explain physical mechanisms of therapies based on applying magnetic fields.

## 1. Introduction

It is well known that a biological system exposed to a physical stimulus is able to detect its presence and to modify its own biological activity depending on the characteristics of the applied stimulus such as mechanic, electric or magnetic. In particular, it has been known since last few decades that electromagnetic fields of weak intensity in the broad interval of frequencies and intensities can cause various biological effects. Such effects can be positive or detrimental. Especially important from the point of view of technical appliances in the modern life and promising from the point of view of medical applications are low and extremely low frequency (ELF) fields which are widely used in various therapies to treat broad class of diseases. In particular, one of such medical applications is based on the therapy, administered with the devices produced by THERESON company. The corresponding method, called Therapeutic Magnetic Resonance TMR™, turned out to show positive results in treatment of diabetic foot disease and vascular ulcers [1]. The method TMR™ consists of exposing patients to low intensity Pulsating Electro-Magnetic Fields (PEMF), at specific patented protected shapes and frequencies of pulses.

In spite of the fact of relatively long history of using pulsating magnetic fields in medicine, little is known about the mechanisms of such therapies. To understand the physical mechanism(s) of the action of magnetic fields on biological systems and to develop a reliable working principle of the pulsating electromagnetic field (PEMF) therapies it is worth to summarize the experimentally observed biological effects caused by electromagnetic fields. In the next section the review is given of the corresponding experimental data. In Section 3 we suggest some mechanisms, which could lead to such biological effects, and summarize them as a possible working principle of PEMFT. Detailed study of some of the mechanisms will be provided elsewhere in the result of our future studies.

## 2. Review of the experimental data

From the very beginning it is worth to stress that the analysis of the literature with experimental data of the biological effects of electromagnetic fields of low frequencies shows these data sometimes can be contradictory. For instance, in 1990s many experiments have been performed with *Escherichia coli* exposed to oscillating magnetic fields. Out of 18 papers, cited in review [2], in which the corresponding results have been reported, 6 papers describe 'some effect', using terminology of the author, 6 other papers report the effect of the 'opposite sign', and in the rest experiments no effects have been observed. Qualitatively similar situation was reported about the experiments with some enzymes, bacteria and fungi [2].

To a great extent this is an often situation, when we are dealing with biological systems, whose complex structure and organization are determined by many parameters, due to which living matter is very sensitive to some 'minor' details of the experiments which can be indeed minor details for the conventional physical systems, but can be significant for biological systems. This is especially so in the experiments with magnetic fields in view of the high degree of electromagnetic 'pollution' coming from the surrounding technical appliances, power lines, telecommunication nets, computers, inclusions of magnetic materials in the building compounds, etc. Also an important role belongs to local terrestrial magnetic field and solar electromagnetic activity. All these factors can cause a synergetic effect as well. Above this the results can depend on the phase of the development of biological cells and their synchronization if any, 'history' of the system under study, etc.

It is well established, that non-thermal biological effects of magnetic fields depend not only on the concrete biological system, but depend significantly also on the parameters of the field, including the field intensity, frequency, modulation, duration of the exposition, etc. For instance, in the alternating field of 50 Hz at 3 Gs the rate of *lac* transcription in *E. coli* is suppressed, while at 5.5 Gs it increases. Weak magnetic fields of the intensity 0.8-8.0 Gs at 60 Hz increase transcription of gene *c-myc* in mice and humans [3]. More experimental data on the relation stimuli--effect for various systems are cited in [4, 5].

As one could expect, the biological effects of weak static and oscillating magnetic fields are different and, respectively, the mechanisms of such effects can be different. Static magnetic fields affect mainly the nervous system. In particular, such functions as navigation and hunting [6, 7], biological clock [8] and some other biological processes are based on the recognition of magnetostatic changes, mainly through certain cells in the visual system. These effects we will not consider in the present paper. Very differently, low-frequency weak magnetic fields cause physiological effects and are system-wide: at certain low frequencies magnetic fields can affect all types of cells.

More experimental data will be described in the next sections.

## 3. Possible mechanisms of biological effects

From the theory of electromagnetism we know, that the electric and magnetic components of the electromagnetic field are connected through the Maxwell equations. In many cases, considered above, the magnetic component of the oscillating electromagnetic field is weaker than the electric component. Nevertheless, namely magnetic component can be the main one, which affects the biological system. This is possible because biological tissues contain a significant amount of water, which is known to absorb significantly electrical radiation, while the magnetic component can penetrate deeply in the biological tissues, and can affect them locally deeply inside, not only on the skin at the biologically active points, like it is the case of the electrical component of the oscillating electromagnetic fields. Therefore, in this paper we will consider the influence of the magnetic component of the oscillating electromagnetic fields on biological organisms.

At least three physical mechanisms of biological effects of magnetic fields can be identified, namely molecular, supramolecular and system, depending on the level of organisation, at which such effects takes place. Molecular mechanism involves effects of magnetic fields on ions, radicals, paramagnetic particles with unpaired electrons and spin, molecules, macromolecules. Magnetic fields can induce spin singlet-triplet transitions in such molecules, change their states and reactivity. Supramolecular mechanisms involve effects on membranes, mitochondria, microcrystalls, cell nuclei etc. System mechanism is more complex and is based on the synergetic effect of molecular and supramolecular effects and is manifested on the level of the biological system (endocryne, nerve, etc.). The biological effects via this mechanism are more delayed in time, since they are result of the primary effects of the first two mechansims, which are processed with time by the corresponding system or by the whole organism.

The dependence of the biological effects of oscillating magnetic fields on the frequency indicates the resonant character of such effects. Among such resonant mechanisms the first to be mentioned, is the ion cyclotron resonance. In biological systems there are various ions and groups of ions, which are sensitive to the alternating magnetic fields. Due to the Lorentz force, a free ion of the mass *m* and electric charge *Ze* in a static magnetic field of the intensity *B* moves along the circular trajectory with the angular frequency

$$\omega_c = \frac{Ze}{m} B .  \qquad (1)$$

The electromagnetic oscillating signal of the frequency *f* will therefore resonate with ions which have a mass-to-charge ratio m/Z given by the relation

$$\frac{m}{Z} = \frac{e}{2\pi f} B  \qquad (2)$$

Such phenomenon is called ion cyclotron resonance. In a more general case the circular motion of ions is superimposed with a uniform axial motion, resulting in a helix, or in a more complex trajectories depending on the given 'geometry' of the experiment.

Another 'sensors' of magnetic fields in biological organisms are electrons. It is well established that all metabolic processes in living organisms are accompanied by the transport of electrons [9], e.g., in the redox processes or in photosynthesis. Similarly to (1), electron cyclotron resonance frequency is

$$\omega_{c,e} = \frac{e}{m_e} B \qquad (3)$$

Here $m_e$ is mass of a free electron.

It might be useful to recall that in SI units the elementary charge $e$ is measured in Coulombs. Thus, $e= 1.602\times10\text{-}19$ Coulombs, the mass is measured in kilograms, e.g., $m_e= 9.109\times10^{-31}$ kg, the magnetic field $B$ is measured in Teslas, and the angular frequency ω is measured in radians per second.

In the redox processes in respiration, the transport of electrons takes place along the so-called electron transport chain [9, 10]. Such chain represents a series of macromolecules that transfer electrons from one another via redox reactions, so that each compound plays the role of a donor for a molecule 'on the left' and acceptor for the molecule 'on the right'. Electron transport chains take place also in photosynthesis [9, 10], where the energy is extracted from sunlight via redox reactions, such as oxidation of sugars and cellular respiration. The location of electron transport chains varies for different systems: it is located in inner mitochondrial membrane in eukaryotes, where oxidative phosphorylation with ATP synthase takes place, or in thylakoid membrane of the chloroplast in photosynthetic organisms, or in the cell membrane in bacteria [9, 10].

Some molecules in the electron transport chains, like quinone or cytochrome *cyt-c*, relatively small molecular mass. They are highly soluble and can move relatively easy outside the mitochondrial membrane, carrying electron from a heavy donor to a heavy acceptor. In theoretical studies such electron transport systems are modeled as complexes which include a donor molecule weakly bound to a bridge molecule, which in its turn is weakly bound to an acceptor molecule. The bridge itself can be modeled as some potential barrier through which the electron tunneling takes place (see, e.g., [11] and references therein. In some other studies the bridge is modeled as a molecule with super-exchange electron interaction taken into account [12]

It has been shown that properties of electrons in these systems differ little from properties of free electrons. Therefore, their cyclotron resonance frequency is close to the frequency determined in Eq. (3).

Some other molecules in the electron transport chain, such as NADH-ubiquinone oxidoreductase, flavoproteids, cytochrome c-oxidase *cyt-aa₃* and cytochrome *cyt-bc₁* complex are proteins with large molecular weight, and, thus, they are practically fixed in the corresponding membrane. Qualitatively and quantitatively different situation takes place for electrons, when they are transported through these proteins of large molecular mass. First of all, a significant part of such proteins is in alpha-helical conformation, which is stabilized by relatively weak hydrogen bonds between every fourth peptide group (a group of atoms H-N-C=O), so that along the helix there are three hydrogen-bounded polypeptide chains. The softness of hydrogen bonds and quasi-one-dimensional structure of polypeptide chains suggests a possible significant role of the electron-lattice interaction in them. Indeed, it has been shown (see [13]) that this electron-lattice interaction is relatively strong and that it results in a self-trapping of electrons: electrons, transferred into a protein from a donor molecule, create a local deformation of the protein. Such deformation acts as a potential well, which attracts an electron. As a result, a bound state of an electron and lattice deformation is formed. This state is described by the nonlinear Schroedinger equation for the electron wave-function, $\Psi(x,t)$:

$$i\hbar \frac{\partial \Psi(x,t)}{\partial t} + J \frac{\partial^2 \Psi(x,t)}{\partial t^2} + 2Jg|\Psi(x,t)|^2 \Psi(x,t) = 0. \qquad (4)$$

Here $x$ is the coordinate along the polypeptide chain, $J$ is the electron exchange interaction constant coming from the overlap of electron wave-functions on the neighboring peptide groups, $g$ is the dimensionless nonlinearity constant, which is determined by the electron-lattice coupling constant, $\chi$, and elasticity of the hydrogen bond, $w$, through the relation

$$g = \frac{\chi^2}{2Jw}. \qquad (5)$$

Equation (4) has the so-called soliton solution:

$$\Psi(x,t) = \Psi_s(x,t) \equiv \frac{1}{2}\sqrt{g}\, Sech\left[g(x - x_0 - Vt/a)/2\right]\exp(im_e Vx/\hbar + i\varphi_s(t)), \qquad (6)$$

where $V$ is the velocity of the soliton, $a$ is the lattice constant (distance between the neighboring peptide groups), and $\varphi_s(t)$ is the time-depending phase of the soliton.

The corresponding deformation of the chain in the soliton state is proportional to the probability of the electron presence in the corresponding place:

$$\rho(x,t) = \frac{\chi}{w(1-s^2)}|\Psi_s(x,t)|^2 \qquad (7).$$

It follows from the Eqs. (5) and (7), that the electron probability, and chain deformation are localized functions with the width of the localization

$$l_s = \frac{\pi a}{g}. \qquad (8)$$

Such solitons, in fact, are a particular case of large polarons in low-dimensional systems, which represent the crossover between small polarons and almost free electrons. They correspond to the ground electron state (the state with the lowest energy) at intermediate values of the electron-lattice coupling and small values of the nonadiabaticity parameter. Namely such conditions are fulfilled in polypeptides (see [13]).

From Eq. (5) and (7) it follows also, that the electron and lattice deformation propagate along the polypeptide chain with constant velocity V from one end to the other one, from a donor to an acceptor molecule, as a coherent localized wave. Such a wave practically does not emit phonons and is exceptionally stable, able to propagate on macroscopic distances in view of the extremely low energy dissipation and the nonlinear nature of it formation.

In view of the binding of an electron with the lattice deformation, effective mass of a soliton is bigger than the mass of a free electron, $m_e$:

$$m_s = m_e(1 + \Delta_s). \qquad (9)$$

Therefore, the soliton cyclotron resonance frequency is different from the frequency of a free electron:

$$\omega_{c,s} = \frac{e}{m_s} B. \qquad (10)$$

In weak magnetic fields electrons can have arbitrary orientation of their spins, and in such a case bound bisolitons in singlet states correspond to the ground electron state [14]. Such bisoliton represents two extra electrons with antiparallel spins that are localized within the same deformational potential well. Effective mass of a bisoliton is bigger than the sum of masses of two free solitons:

$$m_{bs} = 2m_e(1 + \Delta_{bs}) > 2m_s, \qquad (11)$$

and the corresponding bisoliton cyclotron resonance frequency is given by the relation

$$\omega_{c,bs} = \frac{e}{m_{bs}} B. \qquad (12)$$

The oscillating character of the propagation of soliton and bisoliton with frequencies (10) and (12), respectively, according to the relation (7), is accompanied by the propagation of the local deformation of the polypeptide chain, $\rho(x,t)$, which also is the oscillating function of time. This deformation will excite additional vibrational modes in the polypeptide chain and can change its conformation. Thus, we see that the magnetic field affects the electrosoliton transport, and, therefore, it can affect the redox processes. Indeed, it has been demonstrated experimentally in [15] that electromagnetic induction of protection against oxidative stress takes place. More detailed study is needed for the analysis of the effects of PEMF on the dynamics of solitons and bisolitons, namely, how dynamics of (bi)solitons depends on the frequency, pulse duration and pulse modulation. This will be done in the nearest future.

## 4. Experimental support of the suggested mechanisms

In [16] a table is given of the ion cyclotron resonance frequencies, corresponding to biologically active ions, such as $H^+$, $Li^+$, $Mg^{2+}$, $H_3O^+$, $Ca^{2+}$, $Zn^+$, $K^+$, $arg^{2+}$, *asn*$^+$, *glu*$^+$, *tyr*$^+$. Indeed, it has been shown that ELF EMFs are able to trigger ion and molecular cyclotron resonance phenomena in living systems – see [16, 17] and references therein.

In numerous experimental studies it has been shown, that lymphocytes, fibroblasts, leukemic cells, epithelial, cardiac stem cells, bone cells (osteoblasts), pinealocytes, thymocytes, liver cells (hepatocytes), and salivary gland cells of animals and humans are electromagnetically sensitive. One of the cell structures, able to receive the applied signal, has been identifid to be cell membrane [18]. It has been demonstrated that electric or magnetic filds can affect membrane functions not only by a local effect on ion fluxes or ligand binding, but also by altering the distribution and aggregation of the intramembrane proteins [19]. Among such proteins there are different specialised molecules, such as receptors, enzymes, ion channels, integrins that are essential for many fundamental functions mainly related to signal transduction and cell adhesion. In particular, the influence of electromagnetic exposure on ligand binding to hydrophobic receptor proteins is a possible interaction mechanism [20]. Indeed, studies reveal a modulation in the activity of *adenylyl cyclase* by coupling with specific receptor sites in the membrane surface after the exposure to magnetic and electric fields [21].

Recently, it has been demonstrated that PEMFs mediate the modulation of gene transcription [22]. Some experimental studies show and that chronic exposure to PEMFs may alter human cardiac rhythm, it may enhance the effects during surgery, transplantation or heart attack in humans [23]. The predominant effect of PEMF is also shown on the different phases of bone repair, in particular, it has a positive effect on the repair process [24, 25, 26].

The stimulation of repair processes in clinical practice using the effectiveness of PEMF stimulation for enhancement of bone healing has been reported also in [25, 27, 28].

Such physical stimuli are able to trigger a more complex biologic response such as cell proliferation as it is evident from some clinical results [29-31]. Exposure to PEMF induces an increase in the proliferation of human articular chondrocytes suggesting an important role also in cartilagine repair [32]. On the other hand, PEMFs induce programmed cell death in cultured T-cells and determine a decreased T-cell proliferative capacity [33]. It has been shown that repeating pulses of 1-10 Hz increase expression of genes which translate proteins c-*fos* and zip/28 in cortex of rats [34]. Low-frequency (0.1 Hz) high amplitude modulation (3,000-100,000 Hz) magnetic fields cause significant changes in proliferation and differentiation of neuron stem cells in cortex of rats [35].

It has been established that in inflammation massive infiltration of T-lymphocytes, neutrophils and macrophages into the damaged tissue takes place [36]. The presence of $A_{2A}$ adenosine receptors in human neutrophils suggests that adenosine could play an important role in modulating immune and inflammatory processes. Therefore, activation of $A_{2A}$ receptors by PEMFs may have a relevant therapeutic effect [37, 38]. It is known that neutrophils are the most abundant white cells in the peripheral blood and are usually the first cells to arrive at an injured or infected site. Adenosine, interacting with specific receptors on the surface of neutrophils, has been recognized as an endogenous anti-inflammatory agent [39]. The activation of $A_{2A}$ receptors in human neutrophils affects the immune response in cancer, auto-immune and neurodegenerative diseases and decreases inflammatory reactions [40]. Experimental evidence suggests that PEMFs are able to suppress the extravascular oedema during early inflammation [41]. It has also been demonstrated that the complete healing of wounds depends on the presence of $A_{2A}$ adenosine receptor agonists [42]. It has been reported that PEMFs mediate positive effects on a wound healing, controlling the proliferation of inflammatory lymphocytes and resulting in beneficial affects on inflammatory disease [43].

In biological systems there is a whole set of signaling mechanisms, based on various biochemical reactions [9]. Cells as no other known system can convert one type of stimulus into another, using chains of biochemical reactions involving enzymes. Such enzymes are activated by specific molecules, called messengers. In the endocrine system such messengers are epinephrine, insulin, estrogen, etc., other messengers are cAMP (cyclic AMP), cGMP (cyclic guanosine monophosphate), $IP_3$ (inositol triphosphate), nitric oxide (NO), and other.

Many of them can be affected by oscillating magnetic fields. These chains of biochemical reactions are referred to as transduction pathways. Some of the biochemical reactions include up to 15-20 different intermediate stages, some of which take place with participation of ionic radicals [10]. The mostly studied ionic second messenger in cell magneto-sensitivity is the cellular calcium ion, $Ca^{2+}$. Respectively, many experiments have been performed on the effects of magnetic fields on the processes, involving calcium.

As far, as in 1978, Adey, Bawin and Sabbot [44] studying experimentally radiofrequency effects on chick brain, have discovered that calcium transport is profoundly affected when the radiofrequency signal is modulated by specific extremely low frequencies. It has been shown in [45] the exposure to the magnetic field at the low frequency equal to 7.0 Hz at the intensity 9.2 µT, which provide Ca ion cyclotron resonance frequency, initiates differentiation of pituitary corticotrope-derived AtT20 D16V cells. Similarly, differentiation of human LAN-5 neuroblastoma cells can be induced by extremely low frequency magnetic field [46].

Moreover, not only sensitivity to some frequencies has been shown, but it has been reported in [47] that through the ion cyclotron resonances the magnetic field can be useful in the regenerative medicine. Namely, through such resonance mechanism oscillating magnetic field effects on human epithelial cell differentiation and, thus, it can transfer the information in cells [48, 49]. These effects as well as other, described below, are frequency, intensity, dose, and time dependent.

One of the possible physical mechanisms of the bioeffects of weak magnetic fields at Calcium resonance can be determined with the effect of the field on the $Ca^{2+}$-binding proteins. In 1985 A. Liboff has suggested that calcium and potassium ions can be specifically activated by the magnetic field through the ICR effect, which enhances their transport through membrane ion channels, thereby altering signaling mechanisms and cellular function [50]. These signals are mediated in cells by the cytoplasm, in which water is in a structurized state, known as exclusion zones [51, 53], and plays an important role [53]. According to [54], water is able to form coherent domains, which can store the energy of the electromagnetic field and release it at the resonant frequency.

In the last decades, biophysical studies have shown that not only ICR, but also Molecular Cyclotron Ion Resonance can activate some fundamental biological elements (proteins, vitamins, mineral salts) and make them more mobile, allowing to enter more easily through the cellular membrane thus guiding all the biochemical reactions essential for the normal cellular activity.

A combination of static and alternating magnetic fields can be useful for some positive effects [55-59]. Such combination can change the probability of calcium ion transition between different vibrational energy levels, which, in its turn, affects the interaction of the ion with the surrounding ligands. One can expect that such an effect is maximal when the frequency of the alternating field coincides with the cyclotron frequency of calcium ion or with some of its harmonics, and indeed, it has been observed in [45]. Within such a model some quantitative explanation of the main characteristics of experimentally observed effects has been suggested in [60].

An important effect for understanding the mechanisms of the magnetic fields biological effects is magnetic isotope effect: dependence of the biological response of the system to the substitution non-magnetic isotopes to magnetic ones. In particular, the synthesis of adenosinethriphosphate (ATP) by creatinkinase shows significant isotope effect: enzyme with magnetic isotopes $^{25}Mg^{2+}$ produces 2-3 times more ATP molecules, than enzyme with non-magnetic isotope $^{24}Mg^{2+}$ [5]. Similar effect of low frequency electromagnetic fields on $A_{2A}$ adenosine receptors in human neutrophils has been reported in [61]: magnetic fields of 550 and 800 Gs cause significant effects on the synthesis of adenosinethriphosphate (ATP) by creatinkinase.

Another important biological process, sensitive to magnetic field, is DNA transcription by polymerase. Like ATP synthesis, catalysis of DNA transcription involves metal ions (Mg, Zn, ...) and also shows significant magnetic isotope effect for magnetic ions $^{199}Hg$, $^{25}Mg$, $^{67}Zn$, $^{43}Ca$ [62, 63]. In particular, the speed of the DNA synthesis by beta-polymerase depends on the presence of magnetic or non-magnetic Mg isotope in a catalytic site of the enzyme: enzymes with $^{25}Mg^{2+}$ magnetic ions surpress DNA synthesis by 3-5 times as comapring with enymes with nonmagnetic $^{24}Mg^{2+}$ or $^{26}Mg^{2+}$ ions [64]. Similar results are for for beta-polymerase with zink ions: synthesis with enzymes with magnetic isotopes $^{67}Zn^{2+}$ is 2-3 times slower than with non-magnetic $^{64}Zn^{2+}$ [65].

The mechanism of magnetic isotope effect is connected with the interaction of the unpaired electron of cation-radical $Mg^+$ with the nuclei, which results in the change of the electron spin of the pair: spin convertion of the pair from a singlet to a triplet state takes place. Such change of the spin opens a new reaction channel in a radical pair. In other words, the magnetic nuclei controls electron spin of the pair and its reactivity. Indeed, enzymes with magnetic ions $^{25}Mg^{2+}$ can be activated by magnetic field, while enzymes with a nonmagnetic $^{24}Mg^{2+}$ are inhibited [64]. In a similar way external magnetic field can control such reactivity.

The soliton mechanism of the redox processes in respiration and photosynthesis, described briefly in the end of Section 3, is supported by some experimental data as well. In particular, it has been shown that weak magnetic RF and SMFs increase rate of hemoglobin deoxygenation in a cell [66]. The treatment with specific frequencies of electromagnetic waves corresponding to some optimal regime to optimize the redox balance (rH2) and the acidity (pH) of body fluids to restore the cellular metabolism, has been reported in [67]. A number of basic studies confirmed such effects as correction of membrane potential, activation of enzymatic processes, promotion of intra/extracellular ionic balance, enhancement of the biological availability of the fundamental elements in the cellular metabolism, etc., [68]. The preliminary clinical data suggested the significant impact of such fields on cardiovascular parameters (flow mediated dilation) in healthy volunteers, a stronger and quicker antioxidant effect than antioxidant drugs, improvement in muscular coordination and performance through better recruitment of neuromotor units in neuromuscular diseases, increase in body (muscular) mass in unhealthy patients. The enzymatic activation of the basal metabolism and of the fatty acid metabolism in aged rats [67, 68, 69], a significant improvement in the wound closure and bone fractures healing process, improvement of osteogenesis in osteoporosis have also been shown [67, 70, 71].

Catalysis of oxidation of nicotinamide adenine dinucleotide (NADH), which participates in the redox processes, by molecular oxygen, is performed by peroxidase enzyme. This oxidation is an oscillating reaction with the period of oscillations approximately 100 s, during which concentrations of NADH and $O_2$ oscillate. It has been shown that in these reactions both period of oscillations and their amplitude depend in a non-monotonic way on magnetic field in the interval 1000-4000 Gs. Frequency maximum and amplitude minimum are attained in magnetic fields at 1500 Gs [72].

ELF pulsed-gradient magnetic field may be able to inhibit the growth and division of cancer cell and enhance the host cellular immune response. How the low frequency pulsed-gradient magnetic field induces apoptosis of cancer cell and blocks new blood vessel development remains unknown, but it nevertheless has been found that this could be a new method for the treatment of cancer. It has been reported in [73] that ELF pulsed-gradient magnetic field with the maximum intensity of 0.6–2.0 T, gradient of 10–100 T/m, pulse width of 20–200 ms and frequency of 0.16–1.34 Hz treatment of mice can inhibit murine malignant tumor growth. Such magnetic fields induce apoptosis of cancer cells, and arrest neoangiogenesis, preventing a supply developing to the tumour. It has been reported (see [73] and references therein) that pulsed magnetic field (0.8 T, 22 ms, 1 Hz) inhibits the growth of S-180 sarcoma in mice. Fields of these parameters are used to treat patients with middle and late-stage sarcoma disease. The data have shown that nine of 18 cases showed good improvement, and nine were less well inhibited (see [75] and references therein). Another results of effective treatment of 50 cancer cases with a Nd-Fe-B permanent magnet (0.4 T) were reported, according to [73]. In [73] electron microscopic evidences have been demonstrated that ELF pulsed-gradient magnetic field can not only inhibit the growth of S-180 sarcoma in mice, but can also promote their oncolytic ability of host immune cells. The results of these studies have shown that sarcomas from treated mice were smaller and harder than those from controls: sarcomas in the control group were much larger and softer.

Pulsating magnetic fields affect also DNA. Extensive DNA degradation is characteristic in the early stages of apoptosis. Cleavage of the DNA may yield double-stranded fragments with COOH termini as well as single strands. The free C-ends, in DNA can be labeled with DIG-dUTP by terminal deoxynucleotidyl transferase (TdT), with the incorporated nucleotides being detected in a second incubation step with an anti-DIG antibody conjugated with fluorescein. The immuno-complex has an emission wavelength of 523 nm (green light) when excited at 494 nm. Labeled apoptotic cells, counted under fluorescence microscopy, in treated sarcomas were greater on number than in the controls. The phenomenon of immune cells, including lymphocytes and phagocytes, accumulating around cancer cells to destroy and digest them was seen more prominently in sarcomas exposed to the magnetic field [74]. A decrease of DNA content has been observed by Feulgen staining technique which indicated that magnetic field can block DNA replication and mitosis of sarcoma cells. It is found that a decrease of the mitotic phases of carcoma cells is due to exposure.

It has been reported in [75] that a wide variety of challenging musculoskeletal disorders has been treated successfully over the past two decades. The field parameters of therapeutic, pulsed electromagnetic field were designed to induce voltages similar to those produced, normally, during dynamic mechanical deformation of connective tissues. Clinical applications in orthopaedics: treatment of fractures (non-unions and fresh fractures) and spine fusion have been reported in [76]. More than a quarter million patients with chronically un-united fractures have benefitted from this surgically non-invasive method, without risk, discomfort, or the high costs of operative repair. Many of the non-thermal biological responses, at the cellular and sub-cellular levels, have been identified and found appropriate to correct or modify the pathologic processes for which PEMFs have been used. Not only is efficacy supported by these basic studies but by a number of double-blind trials. Specific requirements for field intensity are being defined. The range of treatable illnesses include nerve regeneration, wound healing, graft behavior, diabetes, and myocardial and cerebral ischemia (heart attack and stroke), among other conditions. Preliminary data even suggest possible benefits in controlling malignancy.

## 5. Conclusions.

In conclusion we stress that there are numerous experimental data which show the biological effects of constant and pulsating magnetic fields in the broad interval of frequencies. Moreover, the magnetic fields are widely used in therapies to treat various diseases. Such therapies are mainly phenomenological with respect to the choice of the field frequencies, shapes of pulses, doses, etc., in view of the lack of knowledge of physical mechanisms of the biological effects of magnetic fields. We have demonstrated that there can be several such mechanisms, acting on different levels of the hierarchy of the organization of living organisms, and indicated possibility of the synergism of the biological effects caused by various physical mechanisms.

Based on the experimental data indicating the dependence of the effectiveness of the redox processes on the external magnetic field, we have suggested that one of such mechanisms can be related with the soliton mechanism of charge transport in the redox processes during respiration or photosynthesis. We have shown that within the soliton mechanism of charge transport the magnetic field can cause a hierarchy of changes from the primary effect on the dynamics of electrosolitons, to the changes of the state of macromolecules, to the effects on the rate of respiration, and, finally, to the effect on the whole metabolism of the system.

**Acknowledgment:** The author acknowledges stimulating discussions with E.Fermi, C.Simmi and D.Zanotti from THERESON Company (Italy) and thanks them for sharing the experimental data on TMR™ therapy. This research was done under the partial support of the Fundamental Research grant of the National Academy of Sciences of Ukraine